\documentclass[referee]{cjaa}           

\usepackage{graphicx}                   
\input{epsf.sty}                        
\input{psfig.sty}                       

\newcommand{\msun}{M$_\odot$}

\begin{document}

   \title{ The Quest for Primordial Stellar Populations and 
   the James Webb Space Telescope 
}


   \author{Nino Panagia
      \inst{}\mailto{}
      }
   \offprints{Nino Panagia}                   

   \institute{ESA/Space Telescope Science Institute, 3700 San Martin Drive,
		 Baltimore, MD 21218, USA\\
             \email{panagia@stsci.edu}
          }

\date{Received~~2003 ; accepted~~2003}

\abstract{
The {\it NASA/ESA/CSA} James Webb Space Telescope ~({\it JWST})~ will be
the successor to the Hubble Space Telescope and may be launched as
early as mid-2011. The key scientific goals for ~{\it JWST} are
discovering and understanding the formation of the first stars and 
galaxies, the evolution of galaxies and the production of elements by
stars,  and the process of star and planet formation. Within this
context, we discuss the expected properties  of the first stellar
generations in the Universe. We find that it is possible to discern
truly primordial populations from the next generation of stars by
measuring the metallicity of high-z star forming objects. The very low
background of ~{\it JWST} will enable it to image and study first-light
sources at very high redshifts, whereas its relatively small collecting
area limits its capability in obtaining spectra of z$\sim$10--15
first-light sources to either the bright end of their luminosity
function or to strongly lensed sources. With a suitable investment of
observing time ~{\it JWST} will be able to detect {\it individual}
Population III supernovae, thus identifying the very first stars that
formed in the Universe.
   }

   \authorrunning{N.\ Panagia }            
   \titlerunning{Primordial Stellar Populations and JWST}  


   \maketitle
%
%
\section{Introduction}           
\label{sect:intro}
The James Webb Space Telescope ({\it JWST}), formerly the Next
Generation Space Telescope ({\it NGST}),  is a cooperative program of
the National Aeronautics and Space Administration ({\it NASA}), the
European Space Agency ({\it ESA}) and the Canadian Space Agency ({\it
CSA})  to develop and operate a large, near- and mid-infrared optimized
space telescope by the end of this decade that can build and expand on
the science opened up by the highly successful Hubble Space Telescope
({\it HST}).  Jointly, {\it NASA}, {\it ESA}, and {\it CSA} will build
~{\it JWST}, whose construction is to start in 2004.    

~{\it JWST}  has the goal of understanding the formation of galaxies,
stars, planets and ultimately, life.~{\it JWST} is specifically designed
for discovering and understanding the formation of the first stars and
galaxies, measuring the geometry of the Universe and the distribution
of dark matter, investigating the evolution of galaxies and the
production of elements by stars, and the process of star and planet
formation.

~{\it JWST}~ has been under study since 1995 and is planned to be launched
around 2011, nearly 400 years after Galileo discovered the moons of
Jupiter, over 60 years after Lyman Spitzer proposed space telescopes,
and about 20 years after the launch of {\it HST}. 

~{\it JWST} has been conceived as an 6 m class deployable, radiatively
cooled telescope, optimized for the 1-5$\mu m$ band, with background
limited sensitivity from 0.6 to 10$\mu m$ or beyond, operating for 10
years near the Earth-Sun second Lagrange point (L2), about 1.5
million km from Earth.  It will be a general-purpose observatory,
operated by the STScI for competitively selected observers from the
international astronomy community.

  \begin{figure}
    \begin{center}
      \includegraphics[width=8.5cm]{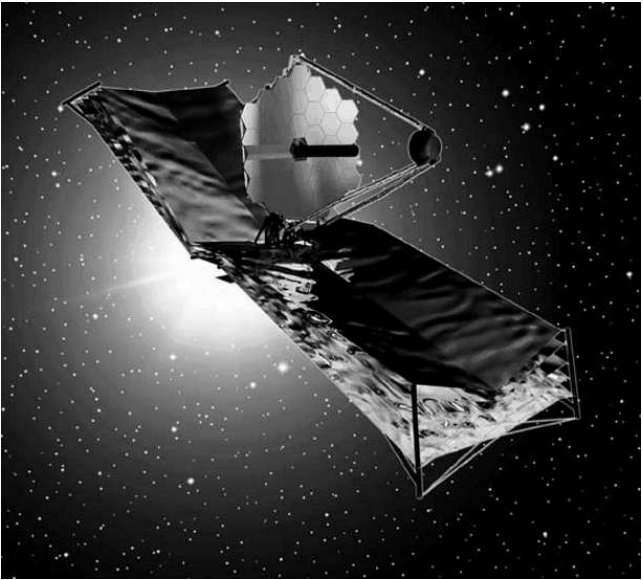}
          \end{center}
  \caption{{The Northrop Grumman design for~{\it JWST}. Easily
recognizable are the Optical Telescope Element (OTE), {\it i.e.}~ the
cylindrical structure between the primary and the secondary mirror, 
and  the multi-layered deployable sunshield. [Copyright \& Credit:
Northrop Grumman  Corporation 2000]}
\label{{fig1}} } 
  \end{figure}
%
%

~{\it JWST} is a unique scientific tool, with excellent angular resolution
(about 0.05 arcsec at 2$\mu m$) over a large field of view (at least
10 arcmin$^2$), deep sensitivity and a low infrared background.  As a
cold space telescope, ~{\it JWST} will achieve far better sensitivities than
ground-based telescopes. Among the advantages of ~{\it JWST} over potentially
competing telescopes are:\\
$\bullet$ ~{\it JWST} will observe with background levels much  lower
than possible from the best sites on Earth: for example, the background
will be one to six orders of magnitude lower than for Mauna Kea,
depending on wavelength, the biggest gain occurring around 5$\mu m$.\\
$\bullet$ ~{\it JWST} will have diffraction limited resolution at 2$\mu
m$, and will achieve much higher Strehl ratios and wider
fields of view than anticipated from ground-based telescopes using
adaptive optics.\\ 
$\bullet$ ~{\it JWST}'s aperture is an order of magnitude larger than
{\it SIRTF}'s, with a factor of 100 better sensitivity.

~{\it JWST} will indeed be able to observe the first generations of stars
and galaxies, including individual starburst regions, protogalactic
fragments, and supernovae out to redshifts of z=5--20. ~{\it JWST} will
resolve individual solar mass stars in nearby galaxies, penetrate
dust-clouds around star-forming regions, and discover thousands of
isolated substellar and Kuiper Belt objects. 

\section{The ~{\it JWST} Mission Concept}

The science goals for ~{\it JWST} require a telescope with high
sensitivity covering the wavelength range from 0.6 to 10$\mu m$, with
capability out to 28 $\mu m$, and with NIR resolution comparable to
that of HST in the optical. Figure 1 shows the observatory that will be
built by the Space Technology division of Northrop Grumman (cleverly
named {\it NGST}) and will include  the Optical Telescope Element
(OTE), the Integrated Science Instruments Module (ISIM) Element, and
the Spacecraft Element (Spacecraft Bus and Sunshield).

The Integrated Science Instrument Module (ISIM) consists of a cryogenic
instrument module integrated with the OTE, and processors, software,
and other electronics located in the Spacecraft Support Module (SMM). 
The instrument suite will include:\\
-- A near IR camera (NIRCam), built by US institutions, covering
0.6-5$\mu m$, critically sampled at 2$\mu m$. The field of about 10
arcmin$^2$ is apportioned over two sub-cameras each covering a field of
2.3'$\times$2.3'.\\
-- A near IR multi-object spectrometer (NIRSpec), provided by ESA, 
with spectral resolutions  1000 and possibly 100, and a spatial
resolution of 100~mas, covering a field of view of 3'$\times$3' and
capable of observing more than 100 objects simultaneously.\\
-- A mid-IR camera/spectrometer (MIRI), built in a 50/50 collaboration
between NASA and European institutions,  covering a field of
2'$\times$2' with a spectral range of 5-28$\mu m$ using a long-slit
cross-dispersed grism with a resolution of 1000.\\
-- A Fine Guidance Sensor (FGS), provided by CSA,  that enables stable
pointing at the milli-arcsecond level and have sensitivity and field of
view to allow  guiding with 95\% probability at any point on the sky. 

The ~{\it JWST} design solve the problem of passively cooling to the
cryogenic temperatures required for NIR and MIR operation by 
{\it (a)} protecting the observatory from the Sun with a multi-layer
shield, 
{\it (b)} using a heliocentric orbit to decrease the Earth's thermal
input, and 
{\it (c)} configuring the telescope to have a large area exposed to
space to improve radiative cooling.

With these general characteristics, ~{\it JWST} will have an enormous
discovery potential both at 0.6-10$\mu m$ and at longer wavelengths. 
In particular,  ~{\it JWST} enjoys a considerable background advantage 
over the ground at all wavelengths, a larger field of view over which
high-resolution images can be obtained and a significant aperture
advantage over {\it SIRTF}.  The shorter times required to reach a
given threshold can translate into larger fields observed (more
targets) and/or greater sensitivities \footnote{Details and updates can
be found at http://www.stsci.edu/jwst/overview/ and linked URLs.}.

\section{Primordial Stars: Expected Properties}
\label{sec:PrimStars}

One of the primary science goals of  ~{\it JWST} is to answer the
question: ``When did galaxies begin to form in the early Universe and
how did they form?" Theorists predict that the formation of galaxies is
a gradual process in which progressively larger, virialized masses,
composed mostly of dark matter, harbor star formation as time elapses.
These dark-matter halos containing stellar populations, then undergo a
process of hierarchical merging and evolution to become the galaxies
that make up the local Universe. In order to understand what are the
earliest building blocks of galaxies like our own, ~{\it JWST} must
detect and identify ``first light" sources, i.e, the emission from the
first objects in the Universe to undergo star formation.

The standard picture is that at zero metallicity the Jeans mass in star
forming clouds is much higher than it is in the local Universe, and,
therefore, the formation of massive stars, say, 100 \msun\/ or higher,
is highly favored. The spectral distributions of these massive stars
are characterized by effective temperatures on the Main Sequence (MS)
around $10^5$~K ({\it e.g.,} Tumlinson \& Shull 2000, Bromm {\it et
al.}~2001, Marigo {\it et al.}~2001).  Due to their temperatures these
stars are very effective in ionizing hydrogen and helium. It should be
noted that zero-metallicity (the so-called population III) stars  of
all masses have essentially the same MS luminosities as, but are much
hotter than their solar metallicity analogues.   Note also that only
stars hotter than about 90,000~K are capable of ionizing He twice in
appreciable quantities, say, more than about 10\% of the total He
content ({\it e.g., } Oliva \& Panagia 1983, Tumlinson \& Shull 2000). 
As a consequence even the most massive population III stars can produce
HeII lines only for a relatively small fraction of their lifetimes,
say, about 1~Myr or about 1/3 of their lifetimes. 

The second generation of stars forming out of pre-enriched material
will probably have different properties because cooling by metal lines
may become a viable mechanism and  stars of lower masses may be
produced (Bromm {\it et al.}\/ 2001). On the other hand, if the
metallicity is lower than about $5\times 10^{-4}$Z$_\odot$,  build up
of H$_2$ due to self-shielding may occur, thus  continuing the
formation of very massive stars (Oh \& Haiman 2002). Thus, it appears
that in the zero-metallicity case one may  expect a very top-heavy
Initial Mass Function (IMF), whereas it is not clear if the second
generation of stars is also top-heavy or follows a normal IMF. 

\section{Primordial HII Regions}

The  high effective temperatures of zero-metallicity stars imply not
only high ionizing photon fluxes for both hydrogen and helium, but
also  low optical  and UV fluxes.  This is because the optical/UV
domains fall  in the Rayleigh-Jeans tail  of the spectrum where the
flux is proportional to the first power of the  effective temperature,
$T_{eff}$,  so that, for equal bolometric luminosity, the actual flux
scales like $T_{eff}^{-3}$.  Therefore, an average increase of
effective temperature of a factor of $\sim2$ will give a reduction of
the optical/UV flux by a factor of $\sim8$. As a result,  one should
expect the rest-frame optical/UV spectrum of a primordial HII regions
to be largely dominated by its nebular emission (both continuum and
lines), so that the best strategy to detect the presence of primordial
stars is to search for the emission from associated HII regions. 

  \begin{figure}
    \begin{center}
      \includegraphics[width=4in]{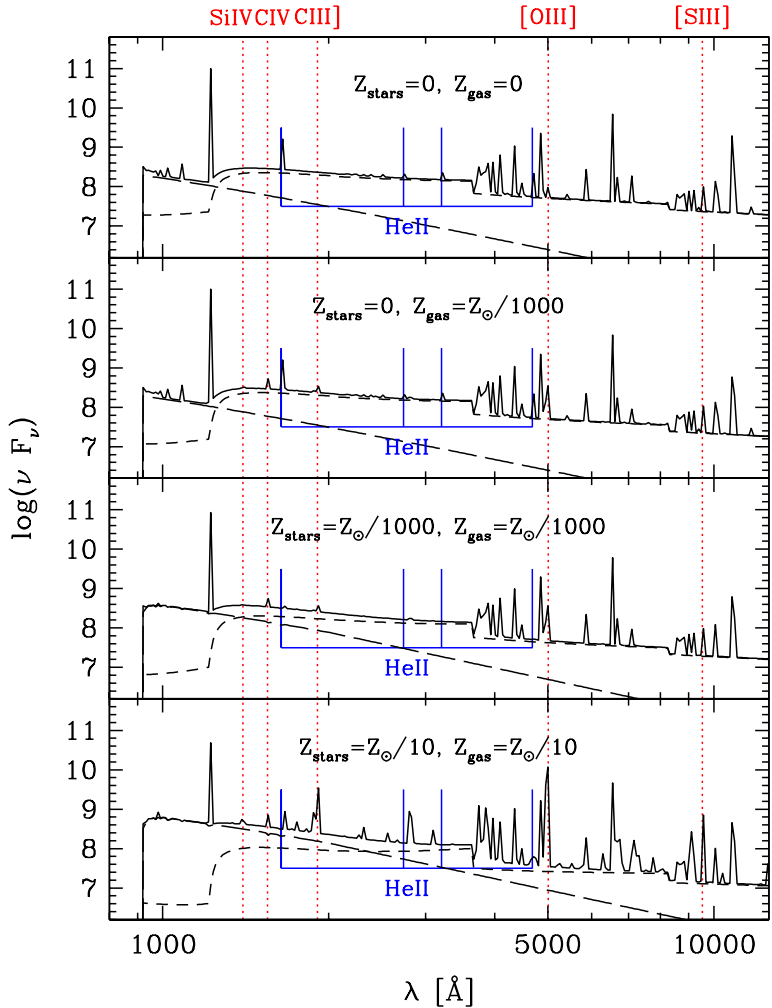}
    \end{center}
  \caption{{The synthetic spectrum of a zero-metallicity HII region
(top panel) is compared to that of HII regions with  various
combinations of stellar and nebular metallicities (lower panels). The
long-dashed and short-dashed lines represent the stellar and nebular
continua, respectively. } 
\label{{fig2}} } 
  \end{figure}
%
%

In Panagia {\it et al.}~(2003, in preparation) we report on our
calculations of the properties of primordial, zero-metallicity HII
regions ({\it e.g.,}~ Figure~2). We find that the electron temperatures
is in excess of 20,000 K and that 45\% of the total luminosity is
converted into the Ly-$\alpha$ line, resulting in a Ly-$\alpha$
equivalent width (EW) of 3000 \AA\/  (Bromm, Kudritzki \& Loeb 2001).
The helium lines are also strong, with the HeII $\lambda$1640 intensity
comparable to that of H$\beta$ (Tumlinson {\it et al.}~2001, Panagia
{\it et al.}~2003, in preparation). An interesting feature of these
models is that the continuum longwards of Ly-$\alpha$ is dominated by
the two-photon nebular continuum. The H$\alpha$/H$\beta$ ratio for
these models is 3.2. Both the red continuum and the high
H$\alpha$/H$\beta$ ratio could be naively (and incorrectly) interpreted
as a consequence of dust extinction even though no dust is present in
these systems.

From the observational point of view one will generally be unable to
measure a zero-metallicity but will usually be able to place an upper
limit to it. When would such an upper limit be indicative that one is
dealing with a population III object? According to Miralda-Escud\'e \&
Rees (1998) a metallicity Z$\simeq10^{-3}Z_\odot$ can be used as a
dividing line between the pre- and post-re-ionization Universe. A
similar value is obtained by considering that the first supernova (SN)
going off in a primordial cloud will pollute it to a metallicity of
$\sim 0.5 \times 10^{-3}Z_\odot$\ (Panagia {\it et al.}~2003, in
preparation). Thus, any object with a metallicity higher than $\sim
10^{-3} Z_\odot$ is not a true first generation object.


  \begin{figure}
    \begin{center}
      \includegraphics[width=7.8cm]{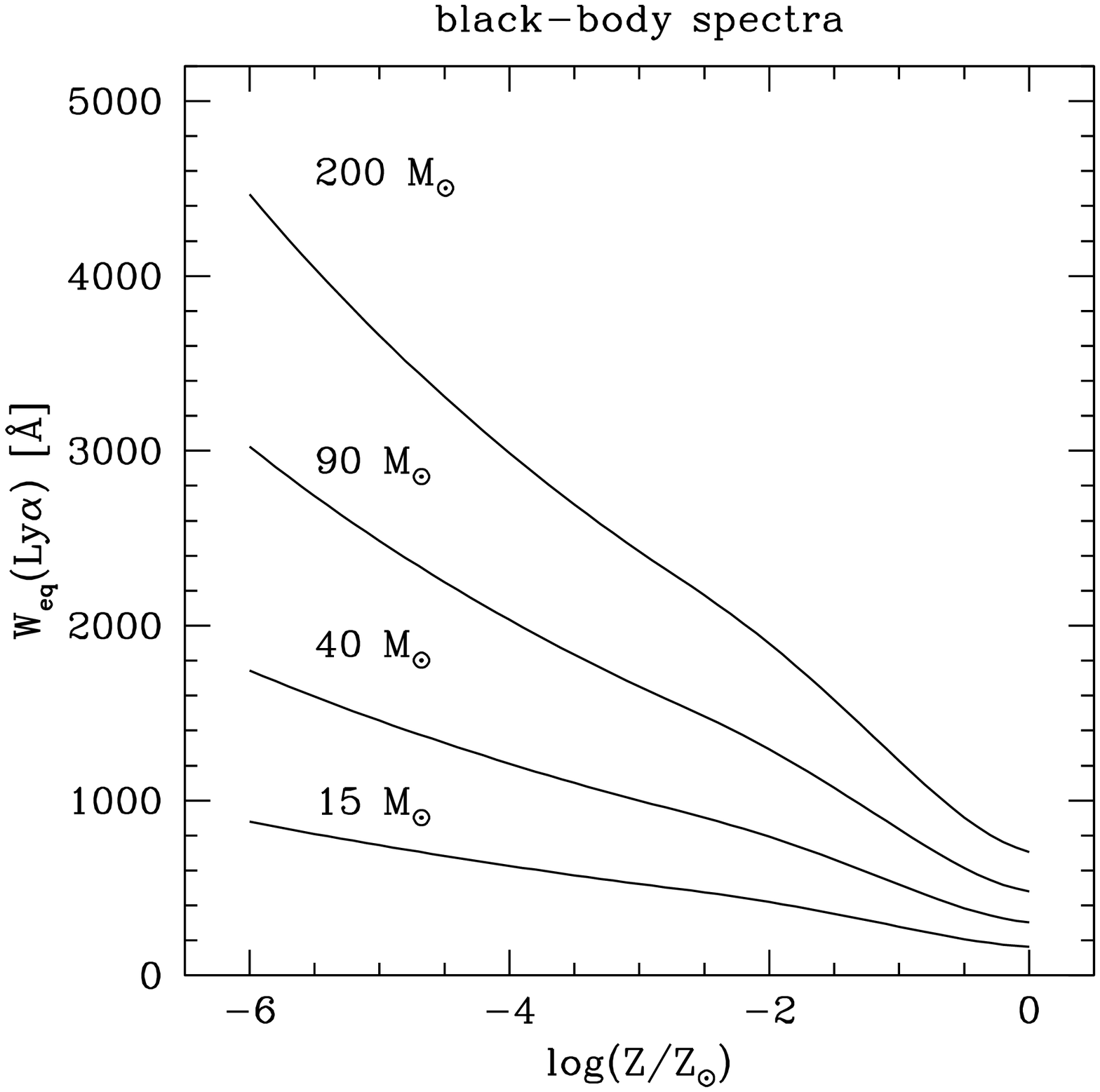}
    \end{center}
  \caption{{Ly-$\alpha$ equivalent widths for HII regions ionized by
stars with a range of masses and metallicities. The results obtained
for black bodies and for stellar atmospheres are very similar to each
other.}
\label{{fig3}} } 
  \end{figure}
   


  \begin{figure}
    \begin{center}
       \includegraphics[width=7.8cm]{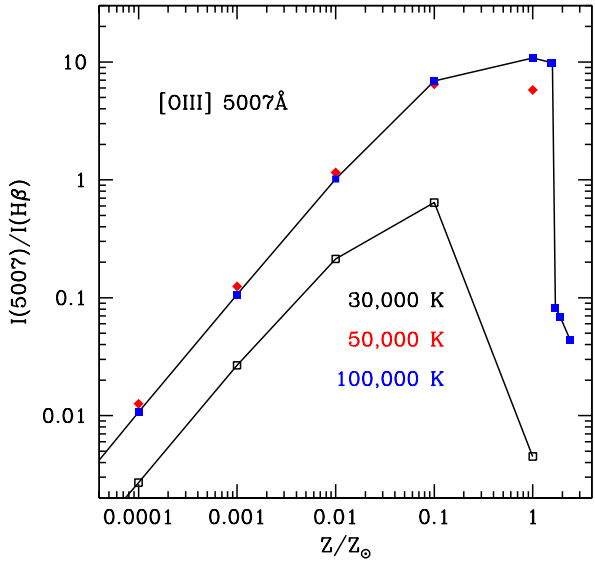}
   \end{center}
  \caption{{The ratio [OIII]$\lambda 5007$ / H$\beta$ is plotted as a
function of metallicity for three different stellar masses: 30,000K
(open squares and bottom line), 50,000K (solid diamonds), and 100,000K
(solid squares and top line).} 
\label{{fig4}} } 
  \end{figure}

\section{Low Metallicity HII Regions}

We have also computed model HII regions for metallicities from three
times solar  down to $10^{-6} Z_\odot$\ (Panagia {\it et al.}~2003,
in preparation).  In Figure~2 the synthetic spectrum of an HII region
with metallicity $10^{-3} Z_\odot$ (third panel from the top) can be
compared to that of  an object with zero metallicity (top panel). The
two are very similar except for a few weak metal lines.  In Figure~3 we
show the Ly-$\alpha$ EWs for HII regions ionized by stars with a range
of stellar masses and metallicities. Values of EW in excess of
1,000\AA\/ are possible already for objects with metallicity $\sim
10^{-3} Z_\odot$. This is particularly interesting given that
Ly-$\alpha$ emitters with large EW have been identified at z=5.6
(Rhoads \& Malhotra 2001).

The metal lines are weak, but some of them can be used as metallicity
tracers. In Figure~4 the intensity ratio of the  [OIII]$\lambda 5007$
line to H$\beta$ is plotted for a range of stellar temperatures and
metallicities. It is apparent that for $Z < 10^{-2} Z_\odot$ this line
ratio traces metallicity linearly. Our reference value $Z = 10^{-3}$
corresponds to a ratio [OIII]/H$\beta$ = 0.1. The weak dependence on
stellar temperature makes sure that this ratio remains a good indicator
of metallicity also when one considers the integrated signal from a
population with a range of stellar masses.


  \begin{figure}
    \begin{center}
      \includegraphics[width=7.8cm]{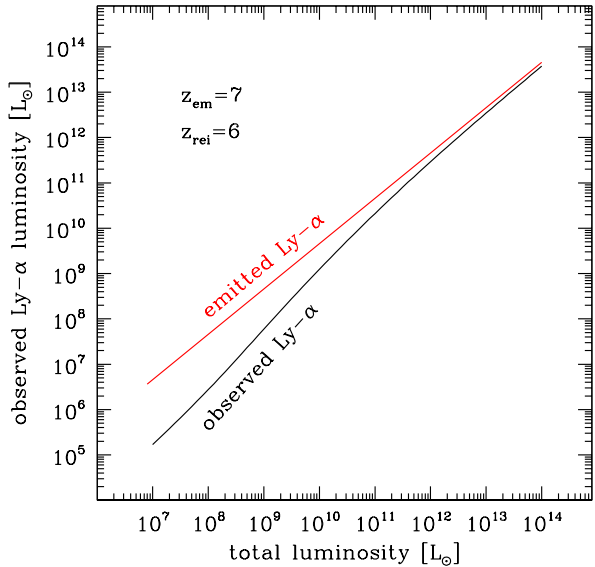}
   \end{center}
  \caption{{Transmitted Ly-$\alpha$ intensity as a function of the object
luminosity. Bright objects ionize their neighborhood and are
able to reduce the Ly-$\alpha$ attenuation.} \
label{{fig5}} } 
  \end{figure}
 

Another difference between zero-metallicity and low-metallicity HII
regions lies in the possibility that the latter may contain dust. For
a $Z=10^{-3} Z_\odot$ HII region dust may absorb up to 30 \% of the
Ly-$\alpha$ line, resulting in roughly 15 \% of the energy being
emitted in the far IR (Panagia {\it et al.}~2003, in preparation).

\section{How to discover and characterize  Primordial HII Regions}

It is natural to wonder whether primordial HII regions will be
observable with the generation of telescopes currently on the drawing
boards. In this section we will focus mostly on the capabilities of the
James Webb Space Telescope.


  \begin{figure}
    \begin{center}
     
 \includegraphics[width=7.8cm]{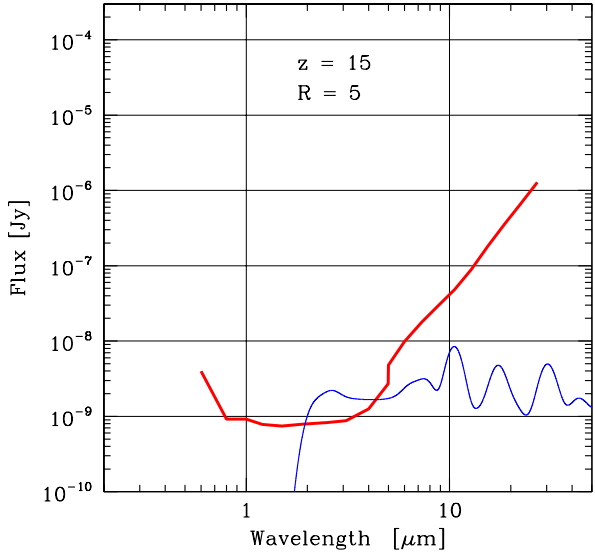}
   \end{center}
  \caption{{Synthetic spectral energy distribution of a Z=$10^{-3} Z_\odot$
starburst object at z=15 containing $10^6$ M$_\odot$ in massive stars
(thin line) compared to the imaging limit of~{\it JWST} at R=5 (thick line).
The~{\it JWST} sensitivity refers to $4\times10^5$ s exposures with S/N=10.} 
\label{{fig6}} } 
  \end{figure}


  \begin{figure}
    \begin{center}
     
\includegraphics[width=7.8cm]{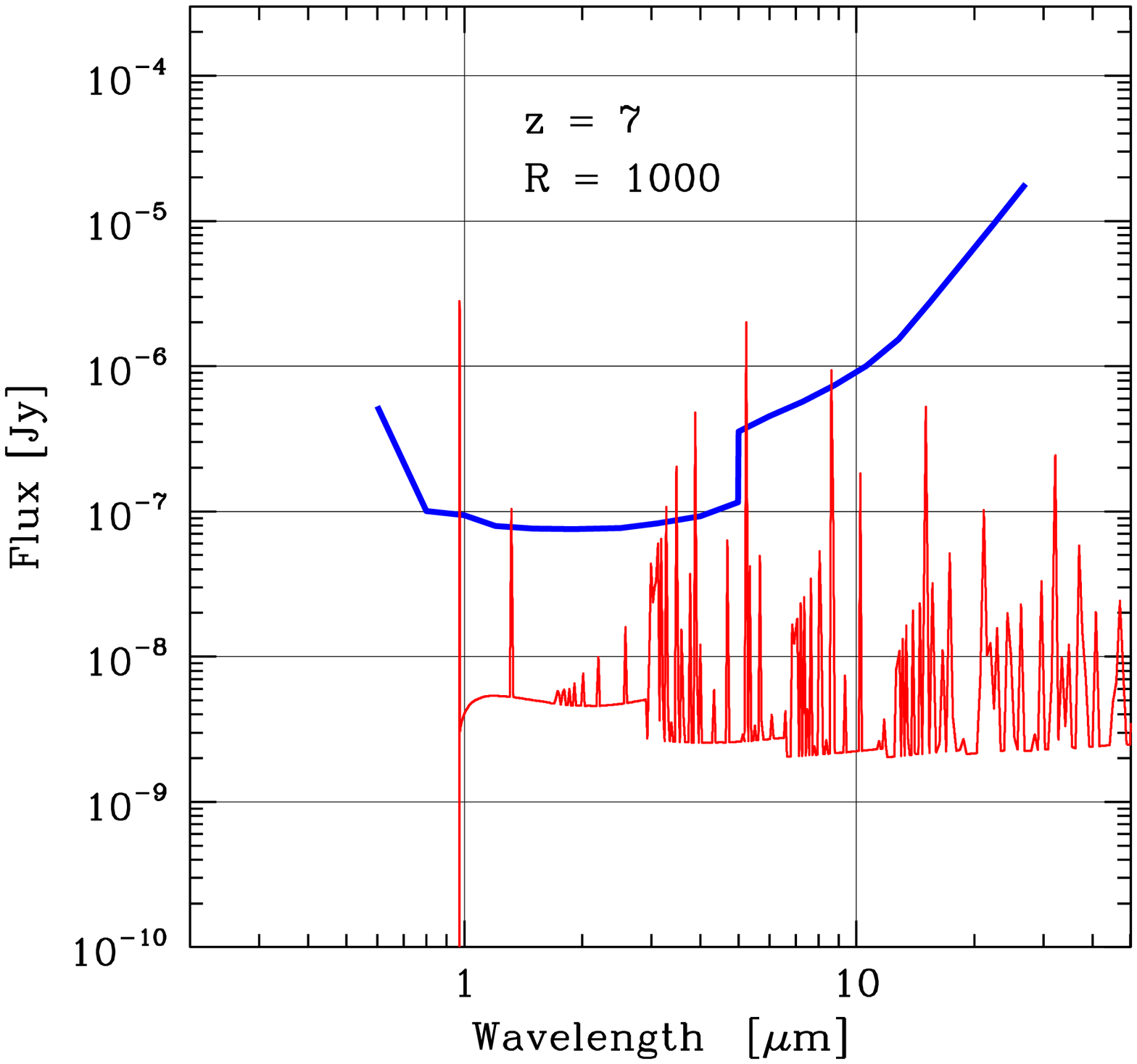}
    \end{center}
  \caption{{Synthetic spectrum of a Z=$10^{-3} Z_\odot$ starburst
object at z=7 containing $10^6$ M$_\odot$ in massive stars (thin line)
compared to the spectroscopic limit of ~{\it JWST} at R=1000 (thick
line). The~{\it JWST} sensitivity refers to $4\times10^5s$  exposures
with S/N=10.} 
\label{{fig7}} } 
  \end{figure}


Before proceeding further we have to include the effect of HI
absorption in the IGM on the Ly-$\alpha$ radiation (Miralda-Escud\'e \&
Rees 1998, Madau \& Rees 2001, Panagia {\it et al.}~2003, in
preparation).  A comparison of the observed vs emitted Ly-$\alpha$
intensities is given in Figure~5.  The transmitted Ly-$\alpha$ flux
depends on the total luminosity of the source since this determines the
radius of the resulting Str\"omgren sphere. A Ly-$\alpha$  luminosity
of $\sim10^{10}$ L$_\odot$ corresponds to $\sim10^6$ M$_\odot$ in
massive stars. In the following we will consider this as our reference
model.

The synthetic spectra, convolved with suitable filter responses  can be
compared to the ~{\it JWST} imaging sensitivity for $4\times10^5$s
exposures (see Figure~6). It is clear that~{\it JWST} will be able to
easily detect such objects. Due to the high background from the
ground,~{\it JWST} will remain superior even to 30m ground based
telescopes for these applications.

The synthetic spectra can also be compared to the~{\it JWST}
spectroscopic sensitivity for $4\times10^5s$ exposures (see Figure 7). 
It appears that while the  Ly-$\alpha$ line can be detected up to
$z\simeq 15-20$, for our reference source only at relatively low
redshifts (z$\sim7$) can~{\it JWST} detect other diagnostics lines lines
such as HeII 1640\AA, and Balmer lines. Determining metallicities is
then limited to either lower redshifts or to brighter sources. 

  \begin{figure}
    \begin{center}
      \includegraphics[width=8.6cm]{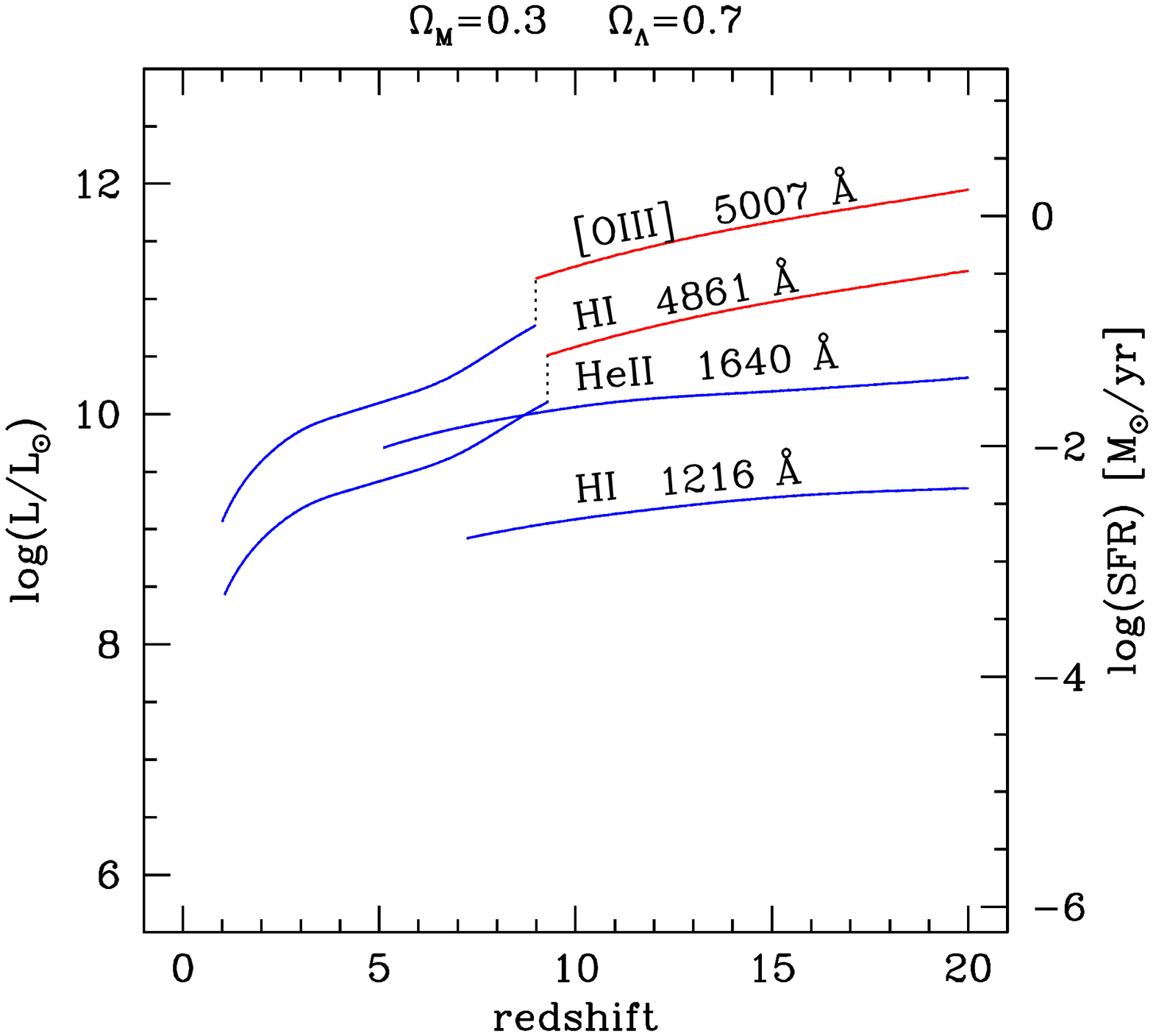}
    \end{center}
  \caption{{Limiting {\it total} luminosity of the ionizing stars
  (left-hand scale) and top-heavy IMF star formation rate (right-hand
  scale) to detect various emission lines using~{\it JWST} spectroscopy,
  with S/N=10 in integrations of 100 hours, as a function of the source
  redshift.} 
  \label{{fig8}} } 
  \end{figure}
%
%

We can reverse the argument and ask ourselves what kind of sources can
~{\it JWST} detect and characterize with spectroscopic observations.  Figure~8
displays, as a function of redshift,  the total luminosity of a
starburst  whose lines can be detected with a S/N=10 adopting an
exposure time of $4\times10^5$s. The loci for Ly-$\alpha$, HeII
1640\AA, H\/$\beta$, and [OIII] 5007\AA\/ are shown. It appears that
Ly-$\alpha$ is readily detectable up to z$\simeq$20, HeII 1640\AA\/ may
also be detected up to high redshifts {\it if} massive stars are indeed
as hot as predicted, whereas ``metallicity" information, {\it i.e.} the
intensity ratio I([OIII])/I(H$\beta$), can be obtained at high
redshifts only for sources that are 10--100 times more massive or that
are 10--100 times  magnified by gravitational lensing.  

\section {Primordial Supernovae}

Even if {\it JWST} cannot detect individual massive Population III
stars, supernova explosions may come to the rescue. We know that local
Universe supernovae (SNe) can be as bright as an entire galaxy ({\it
e.g.}, at maximum light Type Ia supernovae, or SNIa, have
M$_B(SNIa)\simeq$-19.5) and as such, they may be detectable up to large
distances. Practically, SNIa are efficient emitters only at rest frame
wavelengths longer than 3000A, and, therefore, they will be hard to
detect at redshifts higher than z$\simeq$10 (Panagia 2003a,b).
Moreover, the stellar evolution that leads to classical SNIa explosions
is believed to take several hundred million years ({\it e.g.}, Madau,
Della Valle \& Panagia 1998) and, therefore, no type Ia is likely to
occur at redshifts higher than about z$\simeq$8. Type II suypernovae
(SNII) are much more efficient UV emitters but only rarely they are as
bright as a SNIa. As a consequence, they will barely be detected at
redshifts higher than 10 (Panagia 2003a,b) , or, if they are
exceptionally bright (the so-called supernovae of type IIn, \`a la SN
1979C or SN 1998S) they would be rare events.

  \begin{figure}
    \begin{center}
     \includegraphics[width=11cm]{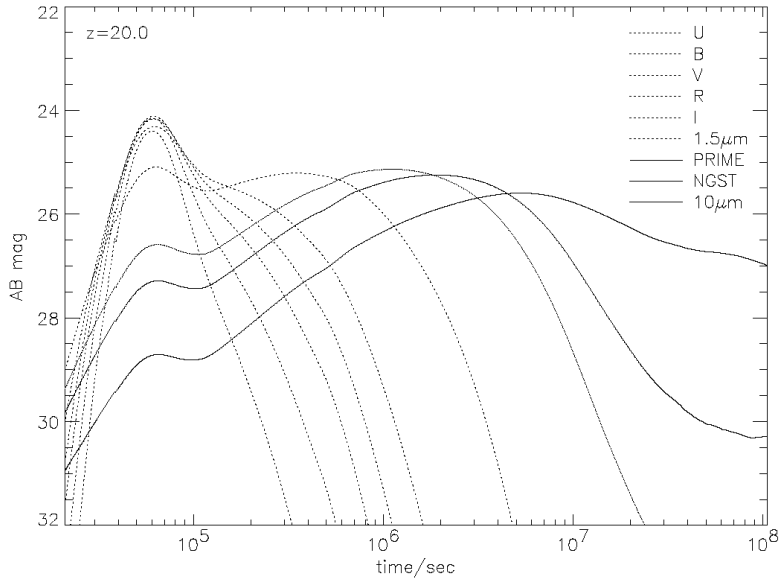}
    \end{center}
  \caption{{Light curves for a pair-creation supernova from a 250 M$_\odot$
star at z=20 (Heger {\it et al.} 2001). Time, wavelengths and magnitudes are
given in observer rest-frame. Wavelengths shorter than the IGM Ly-a
absorption (2.55 $\mu m$) are shown as dotted lines. The curves
labelled as ``PRIME" and ``NGST" correspond to observed wavelengths of 3.5
and 5 $\mu m$, respectively. } 
\label{{fig9}} } 
  \end{figure}

On the other hand, massive population III stars are much more massive
than Pop II or Pop I stars, and the resulting supernovae may have
properties very different from those of local Universe SNe. Heger et al
(2001) have considered the fate of massive stars in conditions of zero
metallicity and have found that stars more massive than 260 M$_\odot$
would directly collapse to a black hole without producing an explosion,
as well as stars with masses in the approximate range 30-140 M$_\odot$.
Below 30 M$_\odot$ the SN explosions would resemble those of type II
supernovae (E$_{kin}\simeq1\times10^{51}$ erg). For stellar progenitors
with masses in the range 140-260 M$_\odot$, Heger et al. find that the
explosions would be caused by a pair-production instability and would
be 3 to 100 times more powerful than than core-collapse (Type II and
Type Ib/c) SNe. As a consequence, even an individual SN may become
bright enough to be detected with~{\it JWST}. In particular, Heger et
al calculate that near maximum light the brightest pair-production SNe
at a redshift of z = 20 may be observed at a flux level of about 100
nJy at 5 $\mu m$, or, correspondingly, be brighter than an AB magnitude
of 26. This high flux is more than 100 times brighter than that of a
typical SNII and, therefore, would be well within the reach of{\it
JWST} observations made with an integration time of a few hours.

While bright supernovae produced by the explosions of primordial Pop
III stars may be ``detectable", do they occur frequently enough to be
found in a systematic search? For a standard cosmology ($\Omega_\Lambda
= 0.7$, $\Omega_m = 0.3$, H$_0=65 ~km~s^{-1}Mpc^{-1}$,
$\Omega_b=0.047$),  and assuming that at z=20 a fraction $10^{-6}$ of
all baryons goes into stars of 250 M$_\odot$,  Heger {\it et al.} (2001)
predict an overall rate of 0.16 events per second over the entire sky,
or about $3.9\times10^{-6}$ events per second per square degree. Since
for these primordial SNe the first peak of the light curve lasts for
about a month (see Figure 9), about a dozen of these supernovae per
square degree should be at the peak of their light curves at any time.
Therefore, by monitoring about 100 NIRCam fields with integration times
of about 10,000 seconds at regular intervals (every few months) for a
year should lead to the discovery of three of these primordial
supernovae. We conclude that, with a significant investment of
observing time (a total of 4,000,000 seconds) and with a little help
from mother Nature (to endorse our theorists views), ~{\it JWST}
will be able to detect individually the very first stars and light
sources in the Universe.

\end{document}